\documentclass[12pt]{article}
\usepackage{epsfig}
\usepackage{amsmath}
\usepackage{hhline}
\usepackage{amssymb}
\usepackage{times}

\newlength{\dinwidth}
\newlength{\dinmargin}
\begin{document}

\begin{titlepage}

\noindent 
DESY 98--075    \hspace*{10.8cm} ISSN 0418--9833
\newline\noindent
June 1998

\vspace*{1cm}
\begin{center}
\boldmath

{\Large 
\bf{Differential (2+1) Jet Event Rates and Determination \\
    of $\alpha_s$ in Deep Inelastic Scattering at HERA}
\unboldmath
 
\vspace*{2.cm}
H1 Collaboration \\
}
 
\end{center}
 
\vspace*{1cm}
 
\begin{abstract}
\noindent
Events with a (2+1) jet topology in deep--inelastic scattering at HERA 
are  studied in the kinematic range $200 <$ Q$^{2}<$ 10\,000 GeV$^{2}$. 
The rate 
of (2+1) jet events has been determined with the modified JADE jet algorithm 
as a function of the jet resolution parameter and is compared with the 
predictions of Monte Carlo models. In addition, the event rate is 
corrected for both hadronization and detector effects and is compared with 
next--to--leading order QCD calculations. A value of the 
strong coupling constant of 
$\alpha_s(M_{Z}^2)$=$0.118 \pm 0.002\,(stat.)\,^{+0.007}_{-0.008}\,(syst.)\,
^{+0.007}_{-0.006}\,(theory)$ 
is extracted. 
The systematic error includes uncertainties in the calorimeter energy 
calibration, in the description of the data by current Monte Carlo 
models, and in the knowledge of the parton densities. The 
theoretical error is dominated by the renormalization scale ambiguity.
\end{abstract}

\vspace{2cm}
 
\centerline{\it Submitted to The European Physical Journal }

\end{titlepage}

\clearpage
\newpage

\noindent
 C.~Adloff$^{34}$,                
 M.~Anderson$^{22}$,              
 V.~Andreev$^{25}$,               
 B.~Andrieu$^{28}$,               
 V.~Arkadov$^{35}$,               
 I.~Ayyaz$^{29}$,                 
 A.~Babaev$^{24}$,                
 J.~B\"ahr$^{35}$,                
 J.~B\'an$^{17}$,                 
 P.~Baranov$^{25}$
% ...

\vspace*{0.5cm}\newline
\noindent
\noindent
% ...
$ ^2$ III. Physikalisches Institut der RWTH, Aachen, Germany$^a$ \\
% ...
$ ^{17}$ Institute of Experimental Physics, Slovak Academy of
          Sciences, Ko\v{s}ice, Slovak Republic$^{f,j}$ \\
% ...
$ ^{22}$ Department of Physics and Astronomy, 
          University of Manchester, Manchester, UK$^b$ \\
% ...
$ ^{24}$ Institute for Theoretical and Experimental Physics,
          Moscow, Russia \\
 $ ^{25}$ Lebedev Physical Institute, Moscow, Russia$^{f,k}$ \\
% ...
$ ^{28}$ LPNHE, Ecole Polytechnique, IN2P3-CNRS, Palaiseau, France \\
 $ ^{29}$ LPNHE, Universit\'{e}s Paris VI and VII, IN2P3-CNRS,
          Paris, France \\
% ...
$ ^{34}$ Fachbereich Physik, Bergische Universit\"at Gesamthochschule
          Wuppertal, Wuppertal, Germany$^a$ \\
 $ ^{35}$ DESY, Institut f\"ur Hochenergiephysik, Zeuthen, Germany$^a$ \\
% ...
$ ^a$ Supported by the Bundesministerium f\"ur Bildung, Wissenschaft,
        Forschung und Technologie, FRG,
        under contract numbers 7AC17P, 7AC47P, 7DO55P, 7HH17I, 7HH27P,
        7HD17P, 7HD27P, 7KI17I, 6MP17I and 7WT87P \\
 $ ^b$ Supported by the UK Particle Physics and Astronomy Research
       Council, and formerly by the UK Science and Engineering Research
       Council \\
% ...
$ ^f$ Supported by the Deutsche Forschungsgemeinschaft \\
% ...
$ ^j$ Supported by VEGA SR grant no. 2/1325/96 \\
 $ ^k$ Supported by Russian Foundation for Basic Researches 
       grant no. 96-02-00019 \\

\clearpage
\newpage

\section{Introduction}
We present a study of events with a (2+1) jet topology and determine 
the strong coupling constant, $\alpha_{s}$, using neutral current 
deep--inelastic scattering (DIS) events recorded with the H1 detector at the 
$ep$ collider HERA in 1994 and 1995. In this period HERA was operated with 
positron and proton beams of 27.5 and 820 GeV energy, respectively, 
corresponding to a centre--of--mass energy of $\sqrt{s} = 300$ GeV.

In the Quark--Parton--Model, neutral current DIS corresponds to the 
interaction of a virtual photon or Z$^0$ boson with a quark in the proton. 
The interaction can be characterized by the two independent variables 
$Q^2$ and $x$ where $Q^2$ is the absolute value of the virtual boson 
4--momentum squared and $x$ is related to the fraction of the proton 
momentum carried by the struck quark. Experimentally,  
events with a (1+1) jet topology are observed. The notation `+1' refers to 
the proton remnant jet. QCD corrections in $O(\alpha_s)$, namely 
QCD--Compton scattering 
($\gamma q \rightarrow q g$) and Boson--Gluon--Fusion 
($\gamma g \rightarrow q\bar{q}$), lead to (2+1) parton final states.
Due to the high centre--of--mass energy at HERA, multi--jet structures 
have been observed clearly \cite{multijet}, and quantitative tests of QCD 
and the 
determination of the strong coupling constant $\alpha_s$ are made 
possible.

Previous jet analyses and determinations of $\alpha_s$ at HERA were based 
on the measurement of $R_{2+1}(Q^2)$, the (2+1) jet event rate as a function 
of $Q^2$ \cite{as-ZEUS-H1}. 
The jets were found by applying the modified JADE jet 
algorithm \cite{JADE} in the laboratory frame for a fixed value of the jet 
resolution parameter. In particular, the measurement of $R_{2+1}(Q^2)$ allows
the dependence of $\alpha_s$ on the scale $Q^2$ to be studied in a single 
experiment. In this analysis a complementary approach is adopted. 
We take events in the range $200 < Q^2 < 10\,000$ GeV$^2$ and then cluster 
measured calorimeter energy depositions with the modified JADE jet algorithm 
until (2+1) jets remain. The minimum mass squared of any pair of the (2+1) 
jet four--vectors, scaled by the hadronic energy squared $W^2$, is the 
variable $y_2$, which we study. 
For a clear (1+1) jet event a small value of $y_2$ is expected whereas any event 
with a larger jet multiplicity must result in a large value.

This is the first measurement of differential jet event rates at HERA 
\cite{Flamm}. 
The presence of a strongly interacting particle in the initial state 
gives rise to considerable differences from the situation in $e^{+}e^{-}$ 
annihilation, where differential jet event rates have been studied in 
much detail \cite{bib-LEP}. The proton remnant, the initial state QCD 
radiation, the large momenta of the produced jets in the direction 
of the incoming proton, and finally the uncertainties in the knowledge of the 
parton content of the proton complicate the measurement. The study of the same 
observable in processes as different as $e^+e^-$ annihilation and $ep$ 
scattering, however, may lead to improved understanding of systematic 
uncertainties in the determination of $\alpha_s$ from hadronic final states 
and provides an important test of QCD.

The analysis consists of the following steps. After the data selection, the 
accuracy with which the data 
are described by the colour dipole Monte Carlo model 
ARIADNE \cite{Ariadne} and the leading--logarithm parton shower model 
LEPTO \cite{Lepto} is studied. The rate of (2+1) jet events is corrected 
for detector acceptance, resolution and inefficiencies as well as for 
hadronization effects. A sophisticated correction procedure is used 
that takes migration effects into account. Next, the parton jet distributions 
of these models are compared qualitatively with next--to--leading order 
(NLO) calculations available in the form of the programs MEPJET \cite{Mepjet} 
and DISENT \cite{Disent}
, in order to 
verify that a jet phase space region has been selected in which the NLO 
calculations can be expected to be a good approximation to the data. 
Finally, the NLO calculations are fitted to the corrected data as a function of 
$\alpha_s$, and the systematic uncertainties 
are evaluated.

\section{The H1 detector}

A detailed description of the H1 detector can be found in \cite{H1-detector}. 
The components most relevant for this analysis are the central tracking 
system, the liquid argon calorimeter, the backward electromagnetic 
calorimeter, and the instrumented iron return yoke. 

The central tracking system consists of several inner and outer drift and 
proportional chambers. It is used in this analysis to determine the 
$ep$ collision point and to aid the identification of the scattered positron. 
The tracking system is surrounded by a large liquid argon sampling 
calorimeter covering a polar angle range of 4$^{\circ}<\theta<154^{\circ}$. 
The polar angle $\theta$ is measured with respect 
to the incoming proton beam which is defined to point in the $+z$ direction. 
The electromagnetic and hadronic sections of the liquid argon calorimeter 
correspond in total to a depth of 4.5 to 8 interaction lengths. The energy 
resolution of the liquid argon calorimeter 
for electrons and hadronic showers is $\sigma/E = 12\%/\sqrt{E({\rm GeV})} 
\oplus 1\%$ 
and $\sigma/E = 50\%/\sqrt{E({\rm GeV})} \oplus 2\%$, respectively
\cite{lar-res}. The absolute energy scale for hadronic energy depositions 
is known to better than 4\%, and that for electromagnetic energy depositions 
to better than 3\%.

Since 1995 the backward region of the H1 detector has been equipped with a 
drift chamber and a lead/scintillating--fibre calorimeter. Its main purpose 
is the detection of electrons at small scattering angles. In addition, 
the timing information it provides allows efficient discrimination  against 
out--of--time 
proton beam related background events at early trigger levels. Before 1995 
the backward region was instrumented with a multi--wire proportional chamber, 
a lead/scintillator electromagnetic calorimeter, and a scintillator array for 
timing measurements.

Outside the calorimeters a large superconducting solenoid provides 
a magnetic field of 1.15~Tesla. The instrumented iron return yoke 
identifies energetic muons and detects leakage of hadronic showers.

\section{Event and cluster selection}

Neutral current DIS events are selected using the 
following criteria. We require a scattered positron candidate to be detected 
within $\theta_{e}$ $<150^{\circ}$ so that it is well contained 
within the acceptance of the liquid argon calorimeter. A cluster of 
contiguous energy depositions in the calorimeter is identified as a  
positron candidate if its energy deposition in 
the electromagnetic calorimeter section 
exceeds 80\% of the cluster energy and if its lateral and longitudinal 
profiles are compatible with those of an electromagnetic shower \cite{lip}. 
In addition, its position 
must be matched to a reconstructed track to better than 
1.7$^{\circ}$ in polar angle and to better than 6$^{\circ}$ in azimuthal 
angle. The $Q^{2}$ range is restricted to $ 200 < Q^{2}<$ 10\,000 GeV$^{2}$ 
where  $Q^{2}$ is determined from the scattered positron energy and 
polar angle. 
The cut $Q^{2} > 200$~GeV$^{2}$ offers several advantages: hadronic final 
state particles are better contained in the detector since they must balance 
the transverse momentum of the scattered positron, which is detected in 
the liquid argon calorimeter at large $Q^2$; 
the range of 
$x$ is implicitly restricted to larger values of $x$ where the parton 
density of the proton is better known and where initial--state QCD 
radiation beyond NLO is suppressed.

The measured $z$ coordinate of the primary event vertex is required to be 
within a distance of 30 cm from the nominal $ep$ collision point. The 
time--of--flight information of the backward scintillator array is required 
not to be 
inconsistent with impact times of particles originating from the $ep$ 
collision point. Both cuts strongly reduce proton beam--related background 
events.

The inelasticity $y = Q^2/s x$, calculated from the scattered positron energy 
and polar angle, is required to be smaller than 0.7. This cut 
corresponds to a polar angle dependent minimum positron energy requirement 
to suppress background from misidentified photoproduction events and to 
reduce the influence of QED radiation. The remaining effects of initial and 
final state QED radiation were studied with DJANGO \cite{Django}. They were 
found to be small and are neglected in the following. Photoproduction and 
beam--related background events are further suppressed by requiring  
30 $< E$--$P_{z}<70$ GeV where $E$ and $P_{z}$ are the summed energy and 
longitudinal momentum components of all reconstructed clusters, each assumed 
to be massless. For NC DIS events $E$--$P_{z}$ is ideally expected to be 
55 GeV, corresponding to twice the positron beam energy. The invariant mass 
squared of the hadronic final state, $W^{2}_{da}$, as 
calculated using the double angle method \cite{doublangl} is required to 
exceed 
5\,000 GeV$^{2}$ to ensure a substantial hadronic activity for jet production.
In addition, we reject events where cosmic muons or beam halo muons crossing 
the detector are identified \cite{cosmics}. 

The events recorded were triggered by the electron trigger of the liquid argon 
calorimeter. The above cuts imply that the energy of the scattered 
positron always exceeds 10 GeV. The average trigger efficiency for the 
selected data sample was found to be larger than 99\% and is independent of 
the hadronic final state. 

With these cuts we obtain a sample of 11\,192 deep--inelastic scattering 
events corresponding to an integrated luminosity of 7 pb$^{-1}$. 
The remaining background from beam--gas collision, photoproduction, cosmic 
muon or halo muon events in this sample is negligible. In particular, the 
fraction of photoproduction events is estimated to be less than 0.5\%. 
The fraction of diffractive events \cite{diffr}, defined as events with a 
calorimeter energy deposition of less than 0.5~GeV in a cone of $15^{\circ}$ 
around the beam direction, is of the order of ~1\%. After the application 
of the jet algorithm, further cuts are applied to select a subsample enriched 
with (2+1) jet events.

In this analysis, hadronic jets are reconstructed from the energy 
depositions in the liquid argon calorimeter and the instrumented iron. 
Clusters that are not well measured  or that are not related to the hadronic 
final state are rejected by the following quality cuts: 
the polar angle, $\theta_{clus}$, of a cluster is required to satisfy 
$\theta_{clus}> 7^{\circ}$ to select clusters that are well within the 
geometrical acceptance of the liquid argon calorimeter, and energy depositions 
in the backward electromagnetic calorimeter are discarded since this has 
limited containment for hadrons. Further requirements of less 
importance are: the energy fraction leaking into the instrumented iron is 
required to not exceed 40\%; hadronic clusters must be separated from the 
positron candidate by an angle greater than 10$^{\circ}$; clusters with an 
angle of larger than 50$^{\circ}$ with respect to their closest neighbouring 
cluster are rejected. This latter cut is imposed to decrease the 
sensitivity to isolated noise contributions or to photons radiated from 
the scattered positron. After these 
selections the average number of accepted clusters per event is 37.8.

\section{Jet algorithms and jet event rate definition }

The jets in a given event are found using the JADE jet algorithm~\cite{JADE}. 
The jet algorithm is applied in the laboratory frame 
to the clusters of the liquid argon calorimeter and the instrumented iron 
satisfying the cuts given in section 3. The algorithm is modified 
compared to the version used in $e^{+}e^{-}$--annihilation in two respects: 
(a) the cluster that is attributed to the scattered positron is removed; 
(b) a massless four--vector is determined and is treated as an additional 
cluster by the jet algorithm to account for the longitudinal component of 
the momentum carried by the proton remnant particles escaping 
through the beam pipe.

The jet algorithm calculates the scaled quantity $m_{ij}^{2}/W^{2}$ 
of pairs of clusters or `proto' jets $i$,~$j$, where $W^{2}$ is the total 
invariant mass squared of all clusters entering the jet algorithm. 
The definition of $m_{ij}^{2}$ is taken to be 
$2E_{i}E_{j}\,(1-\cos\theta_{ij})$. Here $E_{i}$ and $E_{j}$ are the energies 
of the clusters $i$ and $j$, and $\theta_{ij}$ is the angle between them.

In its conventional  
form, the jet algorithm combines the pair of clusters $i$, $j$ with the 
minimum $m_{ij}^{2}/W^{2}$ to be a `proto' jet by adding the four--momenta 
$p_{i}$ and $p_{j}$. This prescription is repeated iteratively for the 
remaining 
clusters and `proto' jets until all possible combinations $i$, $j$ lead 
to $m_{ij}^{2}/W^{2} > y_{cut}$, the jet resolution parameter. In the 
present analysis, however, we use the jet 
algorithm to recombine the accepted clusters iteratively up to the point 
where exactly (2+1) jets remain. The smallest scaled 
jet mass given by any combination of the (2+1) jets is defined to 
be the observable $y_2$. The $y_2$ distribution, $1/N_{DIS} \,\, dn/dy_{2}$, 
where  $N_{DIS}$ is the number of deep--inelastic scattering events 
passing the selection of section 3, corresponds to the differential (2+1) 
jet event rate. 

The same definition of the (2+1) jets and of the variable $y_2$ is 
used for the analysis of the data and of the Monte Carlo events after 
detector simulation. 
In events simulated at the hadron or parton level and in the NLO 
calculations, the jet algorithm is applied to hadron or parton four--momenta, 
respectively.
The polar angle cut of $7^{\circ}$ which is applied for clusters is 
also applied for hadrons, $\theta_{had}$, and partons, $\theta_{par}$. We 
take all components of the `missing momentum' due to this cut into account 
and do not neglect the mass. 

With these definitions we observe that the smallest mass 
$m_{ij}$ of all possible combinations $i,j$ of the (2+1) jets is most likely 
to be obtained by the combination of the two non--remnant jets. The 
fraction of events in which the minimum mass 
is formed by inclusion of the remnant jet is of the order of 15\% for 
both data and NLO calculations.

In addition to the definition given above, we also measure the 
differential (2+1) jet 
rate and determine $\alpha_s$ using the $E$, $E_{0}$ and $P$ variants 
of the JADE algorithm \cite{rec-scheme} without performing a full analysis 
of systematic errors. For these three algorithms $m_{ij}^2$ is defined as $(p_i + p_j)^2$. For the $E$ algorithm, the combined four--momentum is simply 
the sum of the four--momenta $p_i + p_j$. For the $E_0$ algorithm, the 
combined energy is defined as $E_i$ + $E_j$ and the combined momentum is 
$\frac {E_i + E_j}{|\vec{p_i} + \vec{p_j}|} (\vec{p_i} + \vec{p_j})$.
For the $P$ algorithm, the combined momentum is $\vec{p_i} + \vec{p_j}$ and 
the combined energy is $|\vec{p_i} + \vec{p_j}|$. The definition of the 
recombination scheme for the latter two algorithms implies that the 
reconstructed jets are massless. This is not the case for the JADE 
algorithm and its $E$ variant which conserve energy and momentum exactly in 
the recombination procedure.

\section{Description of the data by QCD models}

Before correcting the (2+1) jet event rate for detector and hadronization 
effects, as described in the next section, we study the description of 
the data by the QCD models LEPTO 6.5 and ARIADNE 4.08.
LEPTO is based on the exact first order matrix 
elements followed by higher order radiation approximated by leading 
logarithm parton showers.
In contrast, ARIADNE models the QCD cascade by emitting gluons from a 
chain of radiating colour dipoles. In QCD--Compton events the dipole 
is formed between the struck quark and the proton remnant, and the 
first gluon emission reproduces the first order matrix elements. 
In boson--gluon--fusion events, the quark and the antiquark are 
generated according to the first order matrix elements. Two dipoles are 
formed between each quark and the proton remnant and continue to 
radiate independently. Both LEPTO and ARIADNE use the Lund string 
hadronization model \cite{string}. We used the parameters of LEPTO and 
ARIADNE tuned to reproduce published HERA data \cite{HERA-WS}, in combination 
with the parton density functions of MRSH \cite{MRSH}. 
The generated events were passed through a full simulation of the H1 detector.
For each model an event sample was generated that was $\sim$~6 times larger 
than that of the experimental data. The same event and cluster cuts 
are applied to the simulated events as to the data.

\begin{figure}[h]
\begin{center}
\epsfig{file=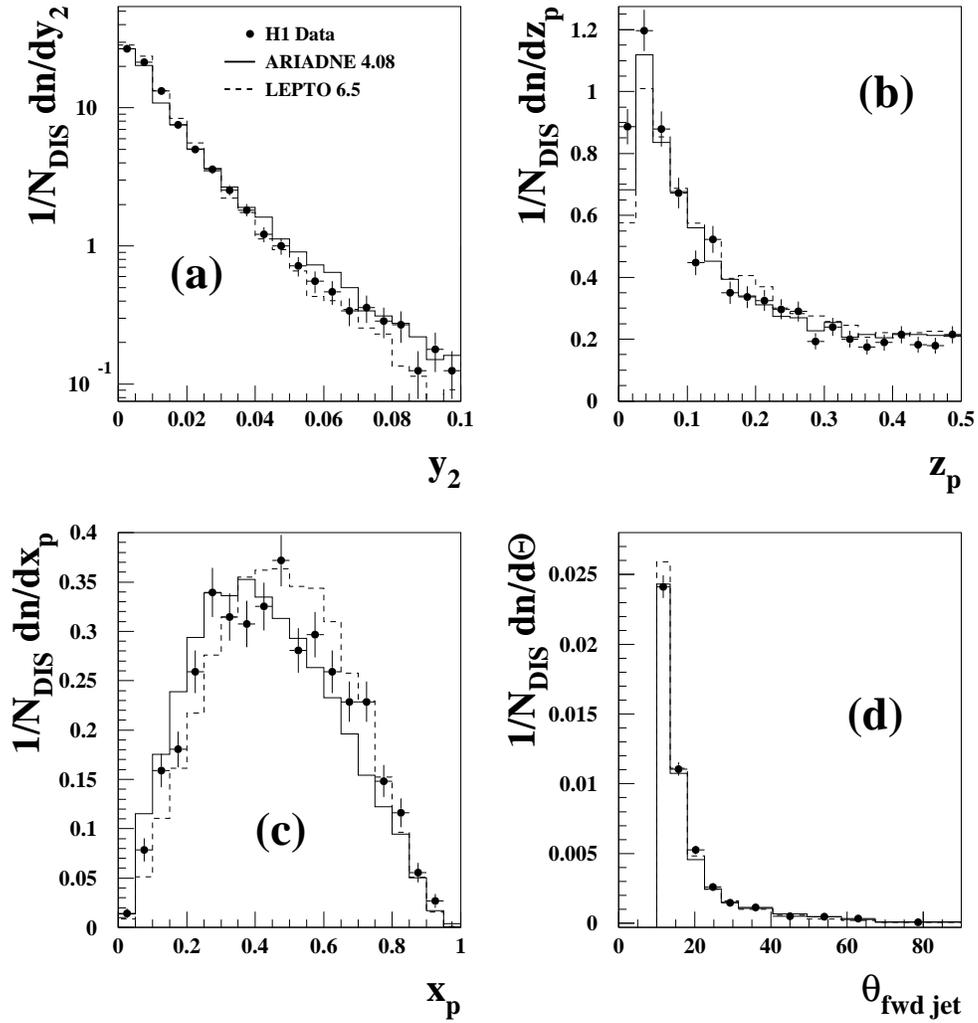,width=14.cm,
clip=}
\end{center}
\caption{Distributions of (a) $y_2$,  (b) $z_p$, (c) $x_p$ and (d) the 
forward jet's polar angle $\theta_{fwd \, jet}$ for uncorrected data 
compared with the prediction of ARIADNE 4.08 and LEPTO 6.5 including full 
detector simulation. For all distributions the (2+1) jet event cuts 
$y_2 > 0.01$ and $10^{\circ} < \theta_{jet} < 145^{\circ}$ are applied 
with the exception of (a) where the cut $y_2 > 0.01$ is omitted. The 
distributions are normalized 
to the number of deep--inelastic events $N_{DIS}$ passing the kinematic cuts. 
The errors are statistical only.}
\label{y2-det}
\end{figure}

In Figure \ref{y2-det} we show the distributions of four representative 
jet variables:
the differential (2+1) jet event rate $y_2$, the variables $z_p$ and 
$x_p$, and the polar angle of the most forward jet. The variable $y_2$ was 
defined above. The definitions of $z_{p}$ and $x_{p}$ are 
$$z_{p} \equiv 
 {\displaystyle\min_{i = 1,2}  E_{i}\,(1 - \cos\theta_{i})}/
{\sum_{i = 1,2} \,\displaystyle E_{i}\,(1 - \cos\theta_{i})}\,\,\,\, {\rm and}
\,\,\,\,
x_{p} \equiv \frac{\displaystyle Q^2}{\displaystyle Q^2 + m_{12}^2} \, ,$$
where $E_{i}$ and $\theta_{i}$ are the energies and polar angles of the two 
non--remnant jets remaining after the clustering of the jet algorithm, and 
$m_{12}$ is the corresponding invariant jet mass calculated without 
neglecting the jets' masses. The variables $x_p$ and $z_p$ 
measure the approach to the (2+1)~$\rightarrow$~(1+1) 
singularities corresponding to the two non--remnant jets becoming 
one jet ($x_p \rightarrow 1$) or as one jet is absorbed into the remnant 
jet ($z_p \rightarrow 0$).

In order to increase the fraction of events with a clear (2+1) jet 
structure, thus enhancing the sensitivity to $\alpha_s$, 
we define a subsample of events with $y_2 > 0.01$. 
To decrease the sensitivity to the modeling of initial--state multi--parton 
emissions and to avoid forward jets which are too close to the proton 
remnant, we require that the non--remnant jets satisfy 
$10^{\circ} < \theta_{jet} < 145^{\circ}$. The requirement 
$\theta_{jet} > 10^{\circ}$ is found, in particular, to improve the 
description of the data by LEPTO. 
After these cuts, the (2+1) jet event sample consists of  2\,235 events.

In Figure \ref{y2-det}(a) the uncorrected $y_2$ data distribution 
is compared with the predictions of LEPTO and ARIADNE. 
The  distribution is normalized to the total number of DIS 
events $N_{DIS}$ selected in section 3. Both models give an acceptable 
description of the data. At large values of $y_{2}$, the distribution 
from ARIADNE tends to be above that of the data while LEPTO is 
systematically low. In Figure \ref{y2-det}(b) and (c) the $z_p$ and $x_p$ 
distribution
are shown for uncorrected data and the models mentioned above. 
ARIADNE roughly describes the measured $z_p$ distribution with the 
exception of the first bin, while LEPTO and data disagree in particular 
in the lowest two $z_p$ bins. The poorest description of the data is 
observed for the $x_p$ 
distribution. ARIADNE approximately reproduces the data in the central part 
of the distribution. It overestimates and underestimates the 
data in the very low and very high $x_p$ region, respectively. LEPTO shows 
the opposite trend. 
Note that the drop of the $z_p$ distribution at the lowest $z_p$ 
bin and the decrease of the $x_p$ distribution at large values of $x_p$ are 
consequences of the cut $y_2 > 0.01$. 
The distribution of the polar angle of the most forward non--remnant jet 
is shown in Figure \ref{y2-det}(d). It is sharply peaked at small angles
and is well described by both models. 

We have studied the accuracy with which the data is described by ARIADNE 
and LEPTO for a wide range of selection criteria in addition to those 
discussed above. Overall ARIADNE gives the better description of the data.
We conclude that the qualitative description of the data is acceptable and 
that a one--dimensional correction of the $y_2$  distribution is possible 
although an improved model description of the data is clearly desirable. 
In the following analysis, we correct the measured $y_2$ distribution with 
ARIADNE and use LEPTO as a consistency check. 
                          
\section{Correction of the data}

\begin{figure}[h]
\begin{center}
\epsfig{file=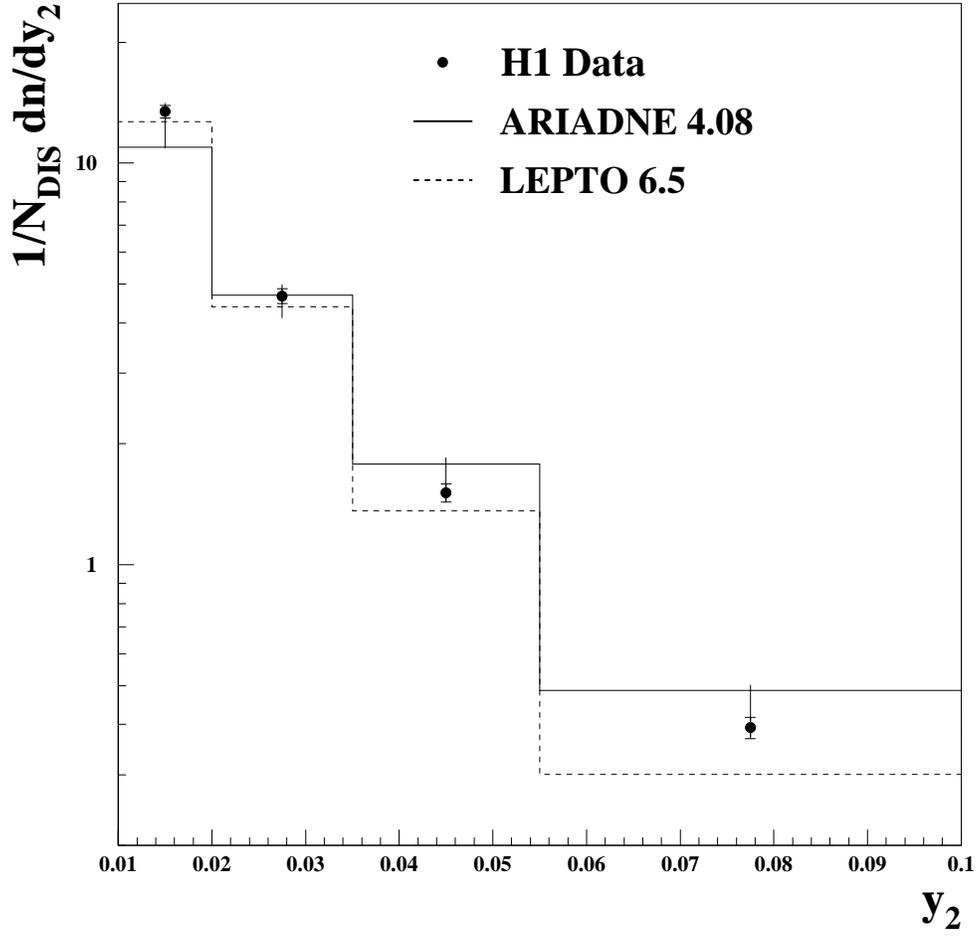,width=14.cm,clip=}
\end{center}
\caption{Distribution of the observable $y_2$ corrected for detector effects 
compared with the prediction of the models ARIADNE 4.08 and LEPTO 6.5. 
As in Figure 1, the cut $10^{\circ} <$ $\theta_{jet}$ $< 145^{\circ}$ is 
applied. The error bars correspond to the statistical and systematic errors 
added in quadrature. The inner error bars give the statistical error only.}
\label{y2-had}
\end{figure}

We correct the measured $y_{2}$ distribution by the method of regularized 
unfolding described in \cite{bib-blobel}. First, we unfold the $y_2$ 
distribution for detector effects only, in order to make direct 
comparisons with QCD model predictions possible. For each simulated ARIADNE 
event, the value 
of $y_2$ is determined by clustering hadrons and simulated calorimeter 
clusters, respectively. 
Then, the $y_2$ distribution calculated from hadrons is reweighted such that 
the  $y_2$ distribution from simulated clusters best fits the 
data. 
The weights are found by means of a log--likelihood method where strongly 
oscillating solutions are suppressed. As result, we obtain four bins of 
a reweighted $y_2$ distribution -- corresponding to unfolded data. 
The unfolded distribution is given in Table \ref{tab-y2}. The quoted 
systematic error 
consists of two contributions added in quadrature: the influence of the 
uncertainty of the absolute hadronic energy scale of the liquid argon 
calorimeter, and the full difference to the $y_2$ distribution unfolded 
with LEPTO instead of ARIADNE. 
The unfolded $y_2$ distribution is shown in Figure~\ref{y2-had} 
together with the predictions of LEPTO and ARIADNE. The statistical 
error is of the order of 5\% but the systematic errors can be larger.
Both models roughly reproduce the data. The prediction of ARIADNE is high 
at large $y_2$, while LEPTO falls too low. These observations are consistent 
with our conclusions from the comparison of the uncorrected data and the 
predictions of LEPTO and ARIADNE in Figure~\ref{y2-det}.

\begin{figure}[h]
\begin{center}
\epsfig{file=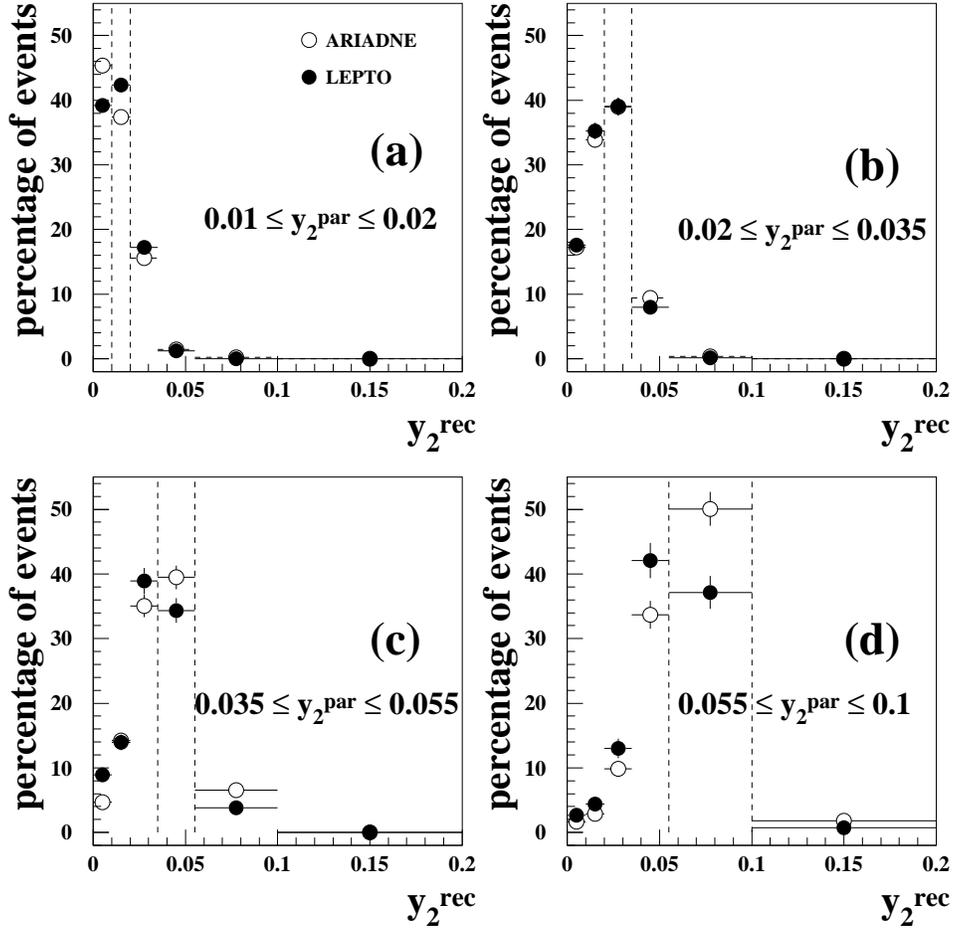,
width=14.cm,clip=}
\end{center}
\caption{ (a) The distribution of $y_{2}^{rec}$ as reconstructed from the 
calorimeter clusters after hadronization and detector simulation from 
the events with $0.01 < y_{2}^{par} < 0.02$ as predicted by
ARIADNE 4.08 (white circles) and LEPTO 6.5 (full circles). The distribution 
is normalized to the number of events with $y_{2}^{par}$ in the range 
$0.01 < y_{2}^{par} < 0.02$. Figures (b), (c) and (d) show the same for 
different ranges of $y_{2}^{par}$ indicated by the legend and by the dashed 
vertical lines.}
\label{y2-corr}
\end{figure}

Next, we unfold both detector and hadronization effects in a one--step 
procedure in order to compare the data to NLO predictions. 
The unfolded distribution is also listed in Table \ref{tab-y2} and is 
discussed in the next section. 
The size of the combined hadronization and detector migration is illustrated 
in Figure~\ref{y2-corr} where the reconstructed $y_{2}^{rec}$ after 
hadronization and detector simulation is compared with the $y_{2}^{par}$ 
found by clustering the partons before hadronization. The bins shown 
correspond to those selected 
for the determination of $\alpha_s$. For both LEPTO and ARIADNE a significant 
correlation is observed between $y_{2}^{rec}$ and $y_{2}^{par}$.
The $y_2$ distribution is systematically shifted to smaller $y_2$ 
values after hadronization and detector simulation, and the 
migrations are sizable. 
This is why a full unfolding procedure is used as opposed to a 
bin--by--bin correction factor method. We study the systematic uncertainty of 
the migrations in detail in section 8 by using alternative QCD models 
for the correction of the data and by varying model parameters. We also 
compared distributions of other jet variables like $x_p$, $z_p$ 
and jet polar angles for partons and for reconstructed clusters after 
hadronization and detector simulation. All the jet variables show clear 
correlations between the different levels.

\section{NLO predictions and determination of {\boldmath $\alpha_{s}$}}

\subsection{NLO QCD programs}
The NLO predictions are calculated with MEPJET, version 1.4 \cite{Mepjet}. 
MEPJET allows arbitrary jet definitions and the application of cuts in terms 
of parton four--momenta. Other programs \cite{Projet} were limited to a 
specific jet algorithm and made approximations in regions of phase space 
relevant for previous $\alpha_s$ analyses \cite{as-ZEUS-H1} that turned out 
to be imprecise \cite{Rome}. MEPJET uses a `phase space slicing' method 
\cite{psslicing} 
to deal with final--state infrared and collinear divergences associated 
with real emissions of partons.  If the invariant mass squared $s$ of 
a pair of partons in a multi--parton state is smaller than a technical 
parameter $s_{min}$, soft and collinear approximations are 
applied to perform the phase space integrations analytically. 
The infrared and collinear divergences thus extracted cancel 
against those from the virtual corrections. If $s$ exceeds $s_{min}$ 
the integrations are done numerically without using explicit 
approximations.  

We run MEPJET with $s_{min}$ set to the 
recommended value of 0.1 GeV$^2$. The statistical precision of the predicted 
$y_2$ distribution is $\sim 1\%$.
As a cross check, we changed $s_{min}$ from 0.1 to 0.05 and 
0.01 GeV$^2$ in MEPJET and observed no significant changes in the 
$y_2$ distributions. Note that our statistical precision at 
$s_{min}$ = 0.01 GeV$^2$ is then reduced to $\sim 2\%$ due to the larger 
fraction of (3+1) parton states treated numerically. 

More recently the 
program DISENT \cite{Disent} became available which uses a different
technique to treat divergences based on a `subtraction' method 
\cite{subtraction} in combination with dipole factorization theorems 
\cite{dipolefac}.
While we use MEPJET for this analysis, we have compared the predictions 
of MEPJET and DISENT version 0.1, which were run with the same value 
of $\Lambda_{\overline{MS}}^{(4)}$ and the same parton densities, for all 
crucial distributions of this analysis and find general agreement at the 
level of a few percent. Looking to the $y_2$ distributions in detail, 
however, we see a significant discrepancy which is of little 
relevance for this analysis and is translated into an error in $\alpha_s$ in 
section 8. To leading order we find the predictions of MEPJET and DISENT to 
be consistent within a fraction of a percent.

\subsection{Comparison of QCD model and NLO predictions}

Before extracting a value of $\alpha_{s}$ from a comparison of corrected 
data and NLO calculations, a region of jet phase space must be identified 
in which NLO predictions can provide a fair description of jet related 
observables. We verify the extent to which this is the case for the above 
selections by comparing NLO jet distributions with the parton jet 
distributions of ARIADNE and LEPTO. The use of QCD model predictions 
rather than corrected data distributions gives reduced statistical error. 
In addition, the comparison of ARIADNE and LEPTO provides interesting 
information on possible ambiguities in the definition of the parton level 
to which the data are corrected. 

\begin{figure}[h]
\begin{center}
\epsfig{file=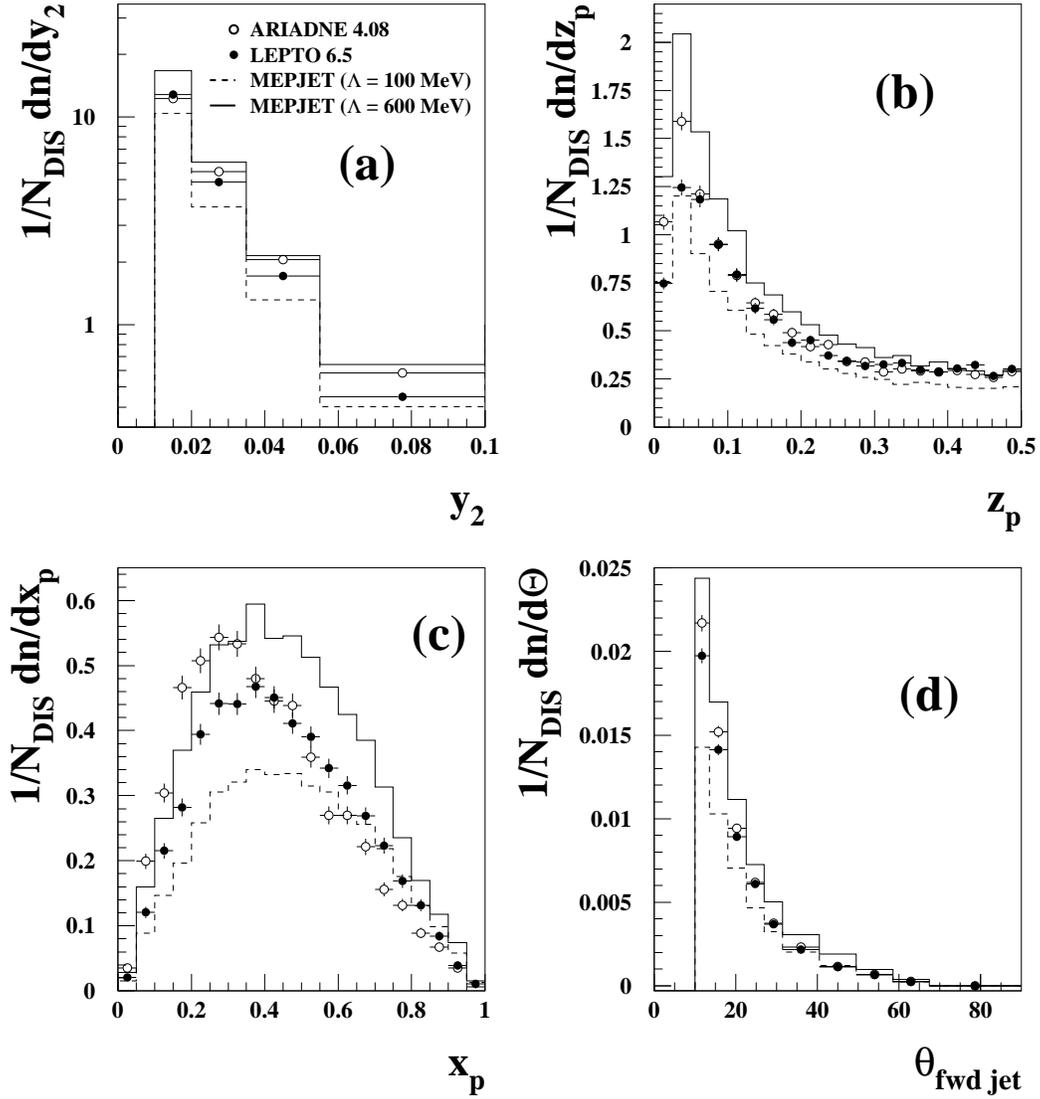,
width=15.cm,clip=}
\end{center} 
\caption{NLO predictions based on MEPJET for the distribution of 
(a) $y_2$, (b) $z_p$, (c) $x_p$ and of (d) $\theta_{fwd\,jet}$ 
compared with parton jet distributions of ARIADNE 4.08 and LEPTO 6.5, 
respectively. The full line corresponds to MEPJET for the extreme 
value of $\Lambda_{\overline{MS}}^{(4)} = 600$~MeV and the dashed 
line to $\Lambda_{\overline{MS}}^{(4)} = 100$~MeV. The cuts $y_2 > 0.01$ and 
$10^{\circ} < \theta_{jet} < 145^{\circ}$ were applied for MEPJET, 
ARIADNE 4.08 and LEPTO 6.5 each.}
\label{zp-dat/NLO}
\end{figure}

In Figure \ref{zp-dat/NLO}(a) the $y_2$ distributions for ARIADNE and 
LEPTO are shown together with NLO calculations for different values of 
$\Lambda_{\overline{MS}}^{(4)}$. 
In order to avoid a dependence of the following study on the value of 
$\alpha_s$ we chose the extreme values of 
$\Lambda_{\overline{MS}}^{(4)} = 100$ MeV and 600 MeV corresponding  
to $\alpha_s(M_{Z}^2)$~=~0.097 and $0.132$, respectively. (Note that 
$\Lambda_{\overline{MS}}^{(4)}$ serves only as a technical steering parameter 
for MEPJET.) The number of flavours used in the calculation is set at five. 
As with ARIADNE and LEPTO, the MRSH parton 
density functions are used in MEPJET. The same cuts on the hadronic 
final state, $y_{2} > 0.01$ and $10^{\circ} < \theta_{jet} < 145^{\circ}$, 
that were applied for Figure 1 are used here. Note that the mean number of 
partons per event with $\theta_{par} > 7^{\circ}$ is 9.7 
for ARIADNE and 10.7 
for LEPTO, whereas in MEPJET at most 
three partons and the proton remnant are produced.

We find that the distributions derived from ARIADNE and LEPTO are in 
qualitative 
agreement, and that their shapes are similar to those of the NLO 
distributions. However at larger values of $y_2$, 
ARIADNE approaches the MEPJET prediction for 
$\Lambda_{\overline{MS}}^{(4)} = 600$ MeV, while LEPTO comes closer to the 
$\Lambda_{\overline{MS}}^{(4)} = 100$ MeV. This trend corresponds to that 
observed from the comparison of data and ARIADNE and LEPTO after detector 
simulation as shown in Figure~1(a).

The predictions of MEPJET and the distribution from ARIADNE for the $z_p$ 
variable, shown in Figure \ref{zp-dat/NLO}(b), are also in fair agreement.
LEPTO falls below ARIADNE at small $z_p$ which is also seen in Figure  
\ref{y2-det}(b). 
A pronounced difference between ARIADNE and NLO is seen in the $x_p$ 
distributions shown in \ref{zp-dat/NLO}(c). This effect is not 
sensitive to changes of the phase space selection criteria and is further 
discussed in section 8. The corresponding prediction 
of LEPTO agrees well with that of the NLO calculations. The distributions 
of the forward jet's polar angle from ARIADNE and LEPTO which are shown 
in Figure \ref{zp-dat/NLO}(d) are well described in shape by QCD in NLO.

We conclude from this comparison that within the phase space region selected 
by the cuts listed above, NLO calculations are expected to provide an adequate 
description of jet production in the data. This statement remains 
qualitative at 
this stage since we do not yet make an assumption on the value of 
$\alpha_{s}$ to be used in the NLO calculations and since we observe 
systematic differences between ARIADNE and LEPTO. 

\subsection{Fit of {\boldmath $\alpha_{s}$}}

\begin{figure}[h]
\begin{center}
\epsfig{file=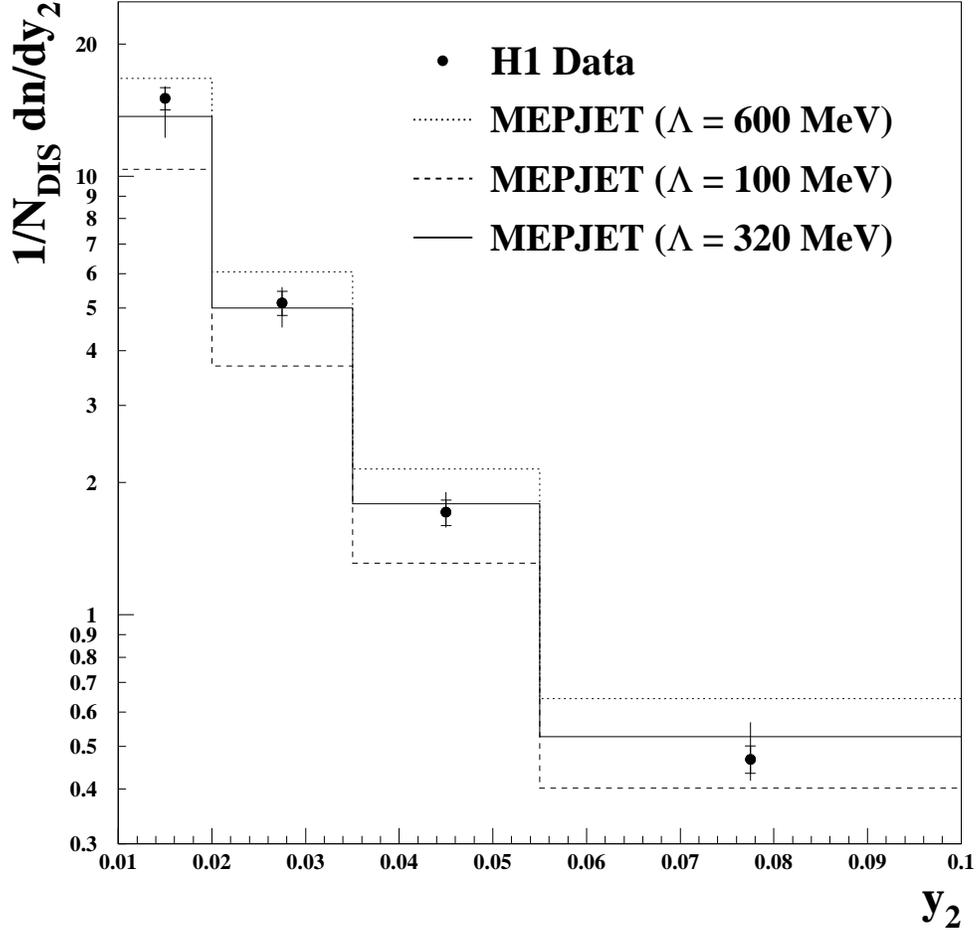,
width=14.cm,clip=}
\end{center}
\caption{Distribution of the differential jet rate $y_{2}$ corrected for 
detector and hadronization effects compared with the NLO prediction of 
MEPJET for $\Lambda_{\overline{MS}}^{(4)} = 600$~MeV (dotted line) and 
$\Lambda_{\overline{MS}}^{(4)} = 100$~MeV (dashed line). The full 
line shows the NLO prediction for the fitted value of $\alpha_s$ 
which corresponds to $\Lambda_{\overline{MS}}^{(4)} = 320$~MeV.
The error bars on the corrected data distribution correspond to the 
statistical and systematic errors added in quadrature. The inner error 
bars give the statistical error only.}
\label{y2-dat/NLO}
\end{figure}

The $y_2$ data distribution corrected for detector and hadronization effects 
is compared with MEPJET in Figure \ref{y2-dat/NLO}. For the first and last 
bin in particular the systematic error (see section 8) is large compared with 
the statistical error and is dominated by the model dependence. The NLO 
predictions of MEPJET for different values of $\Lambda_{\overline{MS}}^{(4)}$ 
are also shown. 

In NLO the differential jet rate is given by the expansion 
$1/\sigma_{DIS} \,\, d\sigma_{2+1}/dy_{2} = A(y_2)\, \alpha_s +  B(y_2) \, 
\alpha_s^2$. From 
the $y_2$ distributions in NLO, obtained by running MEPJET for 
$\Lambda_{\overline{MS}}^{(4)} = 100$ and 600 MeV, we obtain the coefficients 
$A$ and $B$ for the four bins in $y_2$ evaluating $\alpha_s$ at the scale 
$\mu^2 =$ $<Q^2>$, where $<Q^2> \sim$ 620~GeV$^2$ is the mean $Q^2$ of our 
(2+1) jet event sample. The mean $Q^2$ of the entire selected DIS event 
sample is 545 GeV$^2$. 

In order 
to relate $\Lambda_{\overline{MS}}^{(4)}$ to $\Lambda_{\overline{MS}}^{(5)}$ 
and thus to $\alpha_{s}$ at a given scale 
$\mu^2$, we use the following formulae \cite{marc,as-2-loop}

$$  \Lambda_{\overline{MS}}^{(5)} =\Lambda_{\overline{MS}}^{(4)} \,\,\,
    (\Lambda_{\overline{MS}}^{(4)}/m_{b})^{2/23}
       \left[ \ln (m_b^{2}/{\Lambda_{\overline{MS}}^{(4)}} ^{2}) \right]
        ^{-963/13225},
$$
  with $m_b$, the mass of the bottom quark, set to 5 GeV, and 
the two--loop expansion 

$$ \alpha_s(\mu^2) = \frac{4 \pi}{\beta_0 \,\,\ln(\mu^2/
{\Lambda_{\overline{MS}}^{(n_{f})}}^2)} 
                   \left[ 1- \frac{2 \beta_1}
                                  {\beta_0^2} \,\,
                           \frac{\ln \ln(\mu^2/
{\Lambda_{\overline{MS}}^{(n_{f})}}^2)}
                                {\ln(\mu^2/
{\Lambda_{\overline{MS}}^{(n_{f})}}^2)}     \right], $$
\newline
with $ \beta_0 = 11 - \frac{\displaystyle 2}{\displaystyle 3}\,\, n_f$ and 
     $ \beta_1 = 51 - \frac{\displaystyle 19}{\displaystyle 3}\,\, n_f$, and 
     $n_f$ the number of quarks of mass less than~$\mu$, namely $n_f$ = 5 
in our case. The same formulae are used in the MEPJET program. Given $A$ and $B$ 
and the relation of $\Lambda_{\overline{MS}}^{(4)}$ to $\alpha_{s}$, the 
NLO $y_2$ distribution can conveniently be calculated for any value of 
$\Lambda_{\overline{MS}}^{(4)}$.

We perform a minimum $\chi^{2}$ fit ($\chi^{2}/d.o.f. = 6.9/3$) of 
$\Lambda_{\overline{MS}}^{(4)}$ taking into account the statistical 
correlations 
between the bins of the unfolded data distribution. As the result we obtain  
$\Lambda_{\overline{MS}}^{(4)}~=~320 \pm 33$ MeV corresponding to 
$\alpha_s(M_{Z}^2)~=~0.118 \pm 0.002 \,(stat.)$. 
Note that the choice of $\mu^2 =$ $<Q^2>$ for the calculation of the 
coefficients $A$ and $B$ is to some extent arbitrary. It influences the 
value of $A$ and $B$ but not the value of the fitted $\Lambda_{\overline{MS}}^{(4)}$. The NLO prediction corresponding to the fitted value of 
$\Lambda_{\overline{MS}}^{(4)}$ is shown as the full line in Figure 
\ref{y2-dat/NLO}, and a good description of the data is observed.

\section{Determination of systematic errors}

\begin{figure}[h]
\begin{center}
\epsfig{file=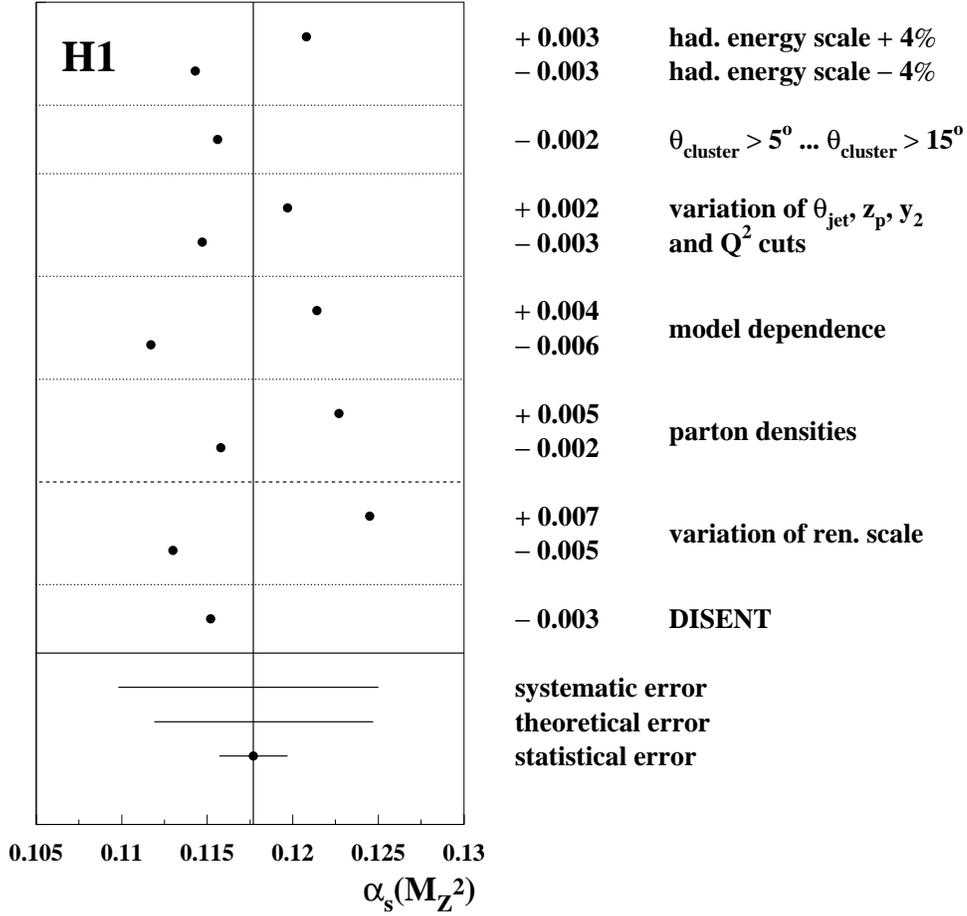,
width=14.cm,clip=}
\end{center}
\caption{List of systematic uncertainties on the fitted value of 
          $\alpha_s$. The horizontal lines separate different 
          classes of uncertainties. The vertical line indicates the 
          central value of $\alpha_s(M_Z)$ resulting from the fit. 
          The black points give the values of $\alpha_s$ that are obtained 
          when each source of sytematic error is varied as described in 
          the text. The uncertainties of the three 
          classes below the dashed horizontal line are combined to give the 
          theoretical error.}
\label{syst-error}
\end{figure}

We study various effects that might influence the result by varying the 
hadronic energy scale of the liquid argon calorimeter, changing the 
experimental cuts, and by using different Monte Carlo models for the data 
correction. We also use alternative parton density functions, measure 
jet rates with 
different variants of the modified JADE jet algorithm, and choose different 
renormalization and factorization scales. The various fitted values of 
$\alpha_{s}$ corresponding to different classes of uncertainties are shown in 
Figure \ref{syst-error}. All values of $\alpha_{s}$ given in the following 
refer to $\alpha_{s}$ at the scale $\mu^2 = M_{Z}^2$. 
\newline

\noindent
{\bf Energy calibration}\newline
The hadronic energy scale of the liquid argon calorimeter is varied by
$\pm 4$\% which leads to a systematic shift in $y_2$. Note that there 
is no fully compensating effect in the ratio $y_2 = m_{ij}^2/W^2$ due 
to the definition of $W$ which includes the `missing momentum' vector. The 
resulting uncertainty in $\alpha_{s}$ is $\pm 0.003$. 

The variation of the electromagnetic energy scale of $\pm 3$\% leads to 
a negligible change in $\alpha_{s}$.
\newline

\noindent
{\bf Polar angle cuts $\mathbf \theta_{clus}/\theta_{par}$}\newline
We vary the minimum value of the cluster acceptance cut $\theta_{clus}$ 
and in parallel the corresponding cut for partons $\theta_{par}$ within a 
range of $5^{\circ} - 15^{\circ}$. The variation of the 
$\theta_{clus}$/$\theta_{par}$ cut checks the quality of the detector 
simulation mostly but also the description of the data in the forward 
detector region where the models are less well tested. It also shows 
the stability of the proton remnant separation by the jet algorithm. 

The additional $\alpha_{s}$ values fitted in this range of cluster cuts 
are slightly lower than the main value, the smallest one differing by 
$-0.002$. We see no indication for a systematic trend in $\alpha_s$ as a 
function of the cut 
value. Without any cluster or parton cut the qualitative agreement between 
NLO and ARIADNE/LEPTO parton distributions deteriorates and stricter phase 
space cuts are needed. As an example we omit the cluster 
or parton cuts as well as the forward jets polar angle cut of 
$10^{\circ}$ but apply the additional event cut $z_p > 0.15$. This reduces
our (2+1) jet event sample by roughly a factor of 2. We obtain an 
$\alpha_{s}$ value of $0.119 \pm 0.003 \,(stat.)$ which is consistent 
with our main result.
\newline

\noindent
{\bf Event selection cuts}\newline
In addition to the $\theta_{clus}$/$\theta_{par}$ cut variation, we study 
the variation or introduction of various event cuts. As before, all cuts are 
applied in parallel to quantities calculated from measured clusters, from 
simulated clusters and partons, and from the partons of the NLO program.
We require $y_2 > 0.02$ instead of $y_2 > 0.01$. We change the polar 
angle jet acceptance cut to $\theta_{jet} > 8^{\circ}$, 
$12^{\circ}$ or $14^{\circ}$. We require $z_p$ to be larger than 0.05, 
0.1 or 0.15. We unfold the differential jet rate for different $Q^2$ ranges 
and vary the minimum $Q^2$ cut from $Q^2 = 200$~GeV$^2$ to $Q^2 = 100$ 
and 250~GeV$^2$. Most of these variations correspond to 
significant changes in the number of events considered. However we find a 
variation of $\alpha_{s}$ of $+0.002$ and $-0.003$ at most. The largest 
change of $-0.003$ is found for the cut $y_2 > 0.02$. Note that the sizes  
of the observed changes in $\alpha_s$ are close to those of our statistical 
error, and that no indication of any systematic trend as a function of a cut 
variation is visible. Thus we regard the analysis as stable with respect to 
the phase space selection.
\newline

\noindent
{\bf Model dependence}\newline
We test the model dependence of the 
result by repeating the analysis using LEPTO for the correction of the data. 
When using cuts identical to those given before, a value of 
$\alpha_{s}$~=~0.116 is obtained. This result is reasonably stable with 
respect to the variation of $\theta_{clus}$, $\theta_{jet}$, $z_p$ and $y_2$ 
cuts, although the observed changes of the determined $\alpha_{s}$ values 
are larger than for the analysis based on ARIADNE. 

Motivated by both the poor agreement of the shape of the $x_p$ distributions 
between ARIADNE and NLO in Figure \ref{zp-dat/NLO}(c) and the relatively large 
differences between the $x_p$ distribution of data and ARIADNE in Figure 
\ref{y2-det}(c), we reweighted ARIADNE events such that the measured $x_p$ 
distribution is reproduced. Effectively, this can be seen as a 
correction to the parton evolution mechanism of ARIADNE. This procedure leads 
to negligible change in the corrected $y_2$ data distribution but we find 
better agreement between ARIADNE and NLO in Figure \ref{zp-dat/NLO}. 
Reweighting ARIADNE in $x_p$ also gives a good description of the $Q^2$ 
dependence of the rate of (2+1) jet events, 
$R_{2+1}(Q^2) = N_{2+1}(Q^2)/N_{DIS}(Q^2)$, where ARIADNE (unweighted) was 
shown to be inferior to LEPTO \cite{r2}.

Possibly large hadronization corrections could fake the radiation of 
hard partons described by perturbative QCD, and could cause systematic
biases in the correction of the data. The uncertainty of the hadronization 
corrections is not directly tested by the comparison of LEPTO and ARIADNE 
since both models use the Lund string hadronization. We thus 
vary the parameters $a$ and $b$ of the Lund fragmentation function 
\cite{string} and the parameter $\sigma_q$, which determines the mean 
$p_t$ of a produced hadron, from their default values $a = 0.3$, 
$b~=~0.58$~GeV$^{-2}$ and $\sigma_q$ = 0.36 GeV, to $a = 0.1$ and 1.0, 
$b = 0.44$ and 0.70 GeV$^{-2}$ and to $\sigma_q$~=~0.25 and 0.45 GeV. We 
derive 
hadronization correction factors for the $y_2$ distributions obtained from 
the events simulated with these sets of parameters. The differences between 
calculated correction factors do not exceed a few percent, and the 
corresponding variations in $\alpha_s$ which we find are at most $+0.002$
 and $-0.002$. In the same manner 
we vary the parameter $Q_0$ of LEPTO which cuts off the  evolution of 
the final state parton shower. Setting $Q_0$ to 4 instead of 1 GeV we observe 
a change of $+0.004$ in $\alpha_{s}$. Setting the corresponding parameter 
for the initial state parton shower from 1.5 to 4 GeV, we observe a change of 
 only $0.001$ in $\alpha_{s}$.
 
We repeat the analysis with the QCD model HERWIG, version 5.8 
\cite{herwig}. HERWIG combines a model for coherent parton shower radiation 
and an additional first order matrix element correction. Hadronization 
follows the cluster fragmentation model \cite{clufrag}. The 
description of the 
data with HERWIG is satisfactory for our purpose although HERWIG does 
not describe the $y_2$ distribution at very small $y_2$ and predicts the 
fraction of (2+1) jet events to be $\sim$~10\% lower than that of the data. 
Unfolding the data with HERWIG leads to a change in the fitted $\alpha_s$ 
value of $-0.006$. As result of the described variation of the models and 
of the model parameters we assess the total model dependence of our 
measurement to be $+0.004$ and $-0.006$. The model dependence represents 
the main source of experimental uncertainty. 
\newline

\noindent
{\bf Parton density functions}\newline
The fit to the experimental $y_2$ distribution is repeated for several 
choices of parton density functions \cite{pdfs} in MEPJET including 
GRV HO (92), CTEQ2pM and CTEQ4M. We find a maximum variation of $+0.005$ 
and of $-0.001$. This dependence is mostly due to the uncertainties in the 
gluon density function. Gluon--initiated processes account for 
$\sim 50\%$ of the (2+1) jet events in our sample.

Since we run MEPJET for values 
of $\Lambda_{\overline{MS}}^{(4)}$ different from those assumed during the 
global fits to deep--inelastic scattering data in which the parton density 
functions were determined, we study the effect of this inconsistency.
This is done using the MRSA', MRSR and CTEQ4A series of parton density 
functions which each combine parton density functions determined on the 
basis of the same experimental data and the same fit procedure but with 
$\Lambda_{\overline{MS}}^{(4)}$ set to different values. From the observed 
change in the fitted value of $\alpha_s$ we estimate the effects of this 
inconsistency to be smaller than $\pm 0.002$. In total, we assign 
an uncertainty of $+0.005$ and $-0.002$ due to the knowledge of the parton 
density functions.
\newline

\noindent
{\bf Different jet algorithms}\newline
In addition to the JADE algorithm, we determine $\alpha_{s}$ with three
related cluster algorithms, namely the $E$, $E_{0}$ and the $P$ algorithms. 
The unfolded differential jet rate distributions are 
given in Table \ref{tab-y2scheme}. 

Comparing the measured $y_2$ distributions 
for the JADE, $E$ and $P$ algorithms, we observe small but statistically 
significant differences. Similar differences are observed for the 
corresponding NLO predictions, which are given in Table 
\ref{tab-y2schememep}, such that the fitted values of $\alpha_s$ do not 
differ much. We obtain $\alpha_s(M_{Z}^2)~=~0.119 \pm 0.002$ and 
$\alpha_s(M_{Z}^2)~=~0.117 \pm 0.002$ for the $E$ and $P$ algorithms, 
respectively. 

The measured $y_2$ distributions for the $E_0$ algorithm is closest to 
that for the JADE algorithm. To next--to--leading order these 
algorithms are identical. This is a consequence of the jet finders' 
definitions and of the fact that in this analysis no cuts on the jets' 
transverse (or longitudinal) momenta are made. The (small) difference 
in the measured $y_2$ 
distributions from these two algorithms may be interpreted as an 
expression of higher order recombination effects which cannot be accounted 
for in $O(\alpha_{s}^2)$ calculations. The value of $\alpha_s$ determined 
with the $E_{0}$ algorithm is $\alpha_s(M_{Z}^2)=0.120 \pm 0.002$. 
The observed differences between the results of the different algorithms 
are small and are not treated as an additional error.
\newline

\noindent
{\bf Renormalization and factorization scale}\newline
In NLO the $y_2$ distribution depends on the choice of the renormalization 
and factorization scales $\mu_{r}^2$ and $\mu_{f}^2$. 
We estimate the renormalization scale dependence by varying $\mu_{r}^2$ from 
$Q^2$ to $1/4~Q^{2}$ and $4~Q^{2}$ in MEPJET and by repeating the 
$\alpha_{s}$ fit. The corresponding uncertainty in $\alpha_{s}$ is 
$+0.007$ and $-0.005$. 
In addition, we use the scalar sum of the transverse momenta of the jets 
in the hadronic centre--of--mass frame as a renormalization scale. This 
corresponds to a considerable difference in the magnitude of the 
renormalization scale given that $Q^2$ is typically about a factor of 20 
larger than the square of a jet's transverse momentum in the hadronic 
centre--of--mass frame.  The corresponding change is close to that 
observed for  $\mu_{r}^2$ = $1/4~Q^{2}$.

In deep--inelastic scattering, 
collinear initial--state divergences are absorbed into redefined parton 
densities introducing the dependence on a factorization scale. We 
estimate the factorization scale dependence by varying $\mu_{f}^2$ from 
$Q^2$ to $1/4~Q^{2}$ and $4~Q^{2}$ as we do for $\mu_{r}^2$, and we 
find the  factorization scale dependence of $\alpha_{s}$ to be $\pm 0.001$.
\newline

\noindent
{\bf DISENT NLO predictions}\newline
As an alternative to the determination of $\alpha_s$ based on MEPJET 
calculations we use DISENT predictions for the fit. The resulting change 
in $\alpha_s$ is $-0.003$. 
\newline

\noindent
{\bf Discussion of higher order effects}\newline
This and similar analyses of $e^{+}e^{-}$ annihilation data
rely on the assumption that the distributions of observables obtained 
from NLO calculations and from the partons in the parton 
shower/colour--dipole models are comparable. There are, however, ambiguities 
in the definition of the parton level of the QCD models used to correct 
the data which can lead to uncertainties in the determined value of $\alpha_s$.
One may argue that the data should not be corrected to the level of the jets 
reconstructed from the final partons before hadronization but to that of the 
jets reconstructed from the partons at an earlier stage of the parton shower 
\cite{opal-reco}. 

We study the dependence on the correction levels in two different ways. 
First, we investigate systematic changes of $y_2$ 
due to subsequent recombinations of partons during the dipole/parton shower 
radiation. We `pre'cluster the final partons of ARIADNE or LEPTO using 
$m_{ij}^2 = 2\min\,(E_{i}^2, E_{j}^2)\,(1-\cos\theta_{ij})$ as used in the 
Durham jet algorithm  \cite{rec-scheme} and then continue clustering using the 
JADE definition. 
The `pre'clustering is stopped when the scaled invariant masses of all 
pairs of parton jets satisfy $m_{ij}^2/W^2 > 0.00005$, where 
$m_{ij}$ is calculated according to the JADE definition. 
This  corresponds  to the point where on average (4+1) parton 
jets remain. We apply the JADE algorithm to these parton jets, continue 
clustering up to (2+1) jets and calculate $y_2$. Comparing the result 
using this procedure with the $y_2$ distribution obtained using 
the JADE algorithm throughout, we see differences of a few percent
for either ARIADNE or LEPTO. This translates into a similar difference 
of a few percent in the result for $\alpha_s$. Larger differences are 
observed, however, when we extend the `pre'clustering further. 

In the same spirit, we also change the value of the parameter $Q_0$ of 
LEPTO, which cuts off the final--state parton shower, 
as a means of looking at an early stage of the parton shower by forcing 
a change in the average number of partons produced.
In contrast to the above variation of $Q_0$ in the context of the model 
dependence, here we are interested in the $y_2$ spectrum of a (variable) 
parton level keeping the hadron level fixed.
We compare the differences in the $y_2$ distributions obtained. Again 
changes of a few percent are observed for e.g. $Q_0 =~3$~GeV instead of 
$Q_0 =~1$~GeV, but differences increase with larger values of $Q_0$. 

Note that the differences in the 
aforementioned definitions of the parton level may partially be due to 
unknown higher order corrections missing in NLO. The interpretation of these 
studies becomes difficult and we have not included such estimates in our 
combined error. Such effects might, in principle, be detected by comparing the shape
 of the distributions from NLO calculations and 
from parton shower/colour--dipole models of Figure \ref{zp-dat/NLO}.
From another viewpoint differences in the measured $\alpha_s$ values using 
different jet recombination procedures, or a large  dependence on the 
renormalization scale could be symptoms of the same difficulty. 
The dependence on the renormalization scale is the dominant uncertainty which is 
included in the systematic error.
This situation with higher order effects and/or the ambiguity of the 
parton level is similar to that
in $e^{+}e^{-}$ annihilation \cite{bib-LEP} where such uncertainties 
turned out to be an important limitation.
\newline

\noindent
{\bf Combined systematic error}\newline
We define the combined systematic error in the following way: 
assuming the errors of the different classes to be largely independent of 
each other, the positive and negative systematic errors of the first 5 
classes are each added in quadrature. Thus we determine 
the systematic error of this analysis to be  $+0.007$ and $-0.008$. An 
additional theoretical error of $+0.007$ and $-0.006$ is obtained 
correspondingly, considering the measured difference between the JADE and 
$E_0$ algorithm, the renormalization scale uncertainty and the 
uncertainty due to observed difference of MEPJET and DISENT.

\section{Summary}

We have presented a measurement of jet related distributions in 
deep--inelastic scattering processes at HERA in the kinematic range 
$ 200 <$ Q$^{2}<$ 10\,000 GeV$^{2}$. The jets are found with the modified 
JADE jet algorithm. The measured 
jet distributions are compared with QCD model expectations and for most 
distributions we find acceptable agreement between the data and the 
models ARIADNE 4.08 and LEPTO 6.5.
Acceptable agreement is also observed for the differential (2+1) jet 
event rate after correcting for detector effects. 

The differential jet rate, corrected for both detector and 
hadronization effects, is compared with NLO QCD calculations in a 
region of jet phase space where the effect of 
higher--order parton emissions not considered in NLO is estimated to be 
small. A fit of the NLO predictions as a function of the strong coupling 
constant $\alpha_s$ is performed which results in
\newline
\begin{center}
$\alpha_s(M_{Z}^2) = 0.118 \pm 0.002 \,(stat.)\,
                           ^{+0.007}_{-0.008} \,(syst.)\,
                          ^{+0.007}_{-0.006} \,(theory)$.
\end{center}
\vspace*{0.5cm} 

A good description of the corrected differential jet rate by the 
next--to--leading order prediction is observed for the fitted value of 
$\alpha_s$. The resulting $\alpha_s$ is compatible with previous 
$\alpha_s$ determinations based on the same observable in $e^+e^-$ 
annihilation \cite{bib-LEP} and with the world average value of $\alpha_s$ 
\cite{as-2-loop} which  provides a direct consistency check of perturbative 
QCD. The same conclusions are reached considering the results 
obtained with the $E$, $E_{0}$ and $P$ variants of the modified JADE algorithm.

The most important uncertainties of the  $\alpha_s$ value determined are 
caused by the as yet limited precision of the data description by current 
QCD Monte Carlo models, by ambiguities in the definition of the parton level 
to which the data are corrected, and by the large renormalization scale 
dependence.

\section*{Acknowledgements} 

We are very grateful to the HERA machine group whose outstanding efforts 
made this experiment possible. We acknowledge the support of the DESY 
technical staff. We appreciate the big effort of the engineers and 
technicians who constructed and maintain the detector. We thank the funding 
agencies for financial support of this experiment. We wish to thank the 
DESY directorate for the support and hospitality extended to the non--DESY 
members of the collaboration.

\clearpage
\newpage

\renewcommand{\arraystretch}{1.5}
\begin{table}[t]  
\begin{center}
\begin{tabular}{|c||c|c|c|}\hline 
 $y_{2}$         &  $1/N_{DIS} \,\, dn/dy_{2}^{had}$  
                 &  $1/N_{DIS} \,\, dn/dy_{2}^{par}$ 
                 & $1/\sigma_{DIS} \,\, d\sigma_{2+1}/dy_{2}$  \\ \hline
 0.010 -- 0.020  & $ 13.43 \pm 0.49 \, ^{+ 0.45}_{-2.49}$ 
                 & $ 15.06 \pm 0.87 \, ^{+ 0.44}_{-2.68} $ 
                 & $ 13.70 \pm 0.15$  \\ 
 0.020 -- 0.035  & $ 4.67 \pm 0.20  \, ^{+ 0.27}_{-0.51}$ 
                 & $ 5.14 \pm 0.32  \, ^{+ 0.31}_{-0.53} $ 
                 & $ 5.01 \pm 0.06$  \\
 0.035 -- 0.055  & $ 1.51 \pm 0.08  \, ^{+ 0.33}_{-0.02}$  
                 & $ 1.71 \pm 0.12  \, ^{+ 0.15}_{-0.05} $ 
                 & $ 1.79 \pm 0.03$ \\
 0.055 -- 0.100  & $ 0.39 \pm 0.02  \, ^{+ 0.11}_{-0.01}$
                 & $ 0.47 \pm 0.03  \, ^{+ 0.09}_{-0.03} $ 
                 & $ 0.53 \pm 0.01$  \\ \hline 
\end{tabular}
\end{center}
\caption{$y_2$ distribution determined with the JADE algorithm corrected for  
         detector effects ($1/N_{DIS} \,\, {\rm d}n/{\rm d}y_{2}^{had}$), 
         corrected for both detector and hadronization effects 
         ($1/N_{DIS} \,\, {\rm d}n/{\rm d}y_{2}^{par}$), and the NLO 
         prediction obtained from MEPJET for $\alpha_s(M_{Z}^2)=0.118$ in 
         combination with the parton density functions MRSH 
         ($1/\sigma_{DIS} \,\, {\rm d}\sigma_{2+1}/{\rm d}y_{2}$). All 
         distributions 
         are determined in the kinematic region defined in section 3 and the 
         cut $10^{\circ} < \theta_{jet} < 145^{\circ}$ is applied for hadron 
         and parton jets, respectively. The first error is statistical, the 
         second systematic. For the NLO calculation only the statistical 
         error is given.}
\label{tab-y2}  
\end{table}

\begin{table}[h!]  
\begin{center}
\begin{tabular}{|c||c|c|c|}\hline 
 $y_{2}^{par}$         & $E$ & $E_{0}$ & $P$ \\ \hline
 0.010 -- 0.020  & $14.97 \pm 0.79$
                 & $15.14 \pm 0.84$ 
                 & $13.00 \pm 0.76$  \\
 0.020 -- 0.035  & $ 6.59 \pm 0.31$  
                 & $ 5.54 \pm 0.32$
                 & $ 4.66 \pm 0.27$  \\
 0.035 -- 0.055  & $ 2.16 \pm 0.13$ 
                 & $ 1.76 \pm 0.12$
                 & $ 1.51 \pm 0.10$  \\
 0.055 -- 0.100  & $ 0.63 \pm 0.05$   
                 & $ 0.49 \pm 0.03$ 
                 & $ 0.42 \pm 0.03$\\ \hline 

\end{tabular}
\end{center}
\caption{$y_2$ distribution corrected for both hadronization and detector 
         effects for the $E$, $E_{0}$ and $P$ algorithms. The error is 
         statistical only.}
\label{tab-y2scheme}  
\end{table}

\begin{table}[h!]  
\begin{center}
\begin{tabular}{|c||c|c|c|}\hline 
 $y_{2}^{NLO}$         & $E$  & $E_{0}$ & $P$ \\ \hline
 0.010 -- 0.020  & $15.59 \pm 0.15$
                 & $14.27 \pm 0.16$ 
                 & $12.39 \pm 0.15$  \\
 0.020 -- 0.035  & $ 5.67 \pm 0.06$  
                 & $ 5.13 \pm 0.06$
                 & $ 4.42 \pm 0.06$  \\
 0.035 -- 0.055  & $ 2.10 \pm 0.03$ 
                 & $ 1.85 \pm 0.03$
                 & $ 1.53 \pm 0.03$  \\
 0.055 -- 0.100  & $ 0.66 \pm 0.01$   
                 & $ 0.53 \pm 0.01$ 
                 & $ 0.46 \pm 0.01$ \\ \hline 
\end{tabular}
\end{center}
\caption{NLO $y_2$ distribution obtained from MEPJET for the $E$, $E_{0}$ and 
         $P$ algorithms for the fitted values of $\alpha_s(M_{Z}^2)~=~0.119$, 
         $0.120$ and $0.117$, respectively, in combination with the parton 
         density functions MRSH.}
\label{tab-y2schememep}  
\end{table}

\clearpage
\newpage

\clearpage
\newpage

\clearpage
\newpage

\clearpage
\newpage

\end{document}